\documentclass[
 reprint,
 bibnotes,
 amsmath,
 amssymb,
 aps,
 prb,
floatfix,
]{revtex4-2}

\usepackage{graphicx}
\usepackage{dcolumn}
\usepackage{bm}
\usepackage{braket}

\usepackage[mathlines]{lineno}

\usepackage{xcolor}

\usepackage{hyperref}

\begin{document}

\preprint{arXiv:}

\title{Charge Tunable Optical Nonlinearity of Moiré Exciton-Polaritons}

\author
{Zhijie Li,$^{1}$ Kok Wee Song,$^{2}$ Jens-Christian Drawer,$^{1}$ Vita Solovyeva,$^{1}$ Martin Esmann,$^{1}$ Niklas Nilius,$^{1}$ Muhammad Sufyan Ramzan,$^{1,3}$ Caterina Cocchi,$^{1,3}$ Falk Eilenberger,$^{4,5}$ Takashi Taniguchi,$^{6}$ Kenji Watanabe,$^{7}$ Atanu Patra,$^{8}$ Subhamoy Sahoo,$^{8}$ Simon Betzold,$^{8}$ Sven Höfling,$^{8}$ Alexander Högele,$^{9,10}$ Oleksandr Kyriienko,$^{11}$ Christian Schneider,$^{1}$ and Bo Han,$^{1,}$}\email{Corresponding author: \\bo.han@uni-oldenburg.de}

\affiliation
{
$^{1}$Institut für Physik, Fakultät V, Carl von Ossietzky Universität Oldenburg, 26129 Oldenburg, Germany\\ 
$^{2}$Department of Physics, Xiamen University Malaysia, 49300 Sepang, Malaysia\\
$^{3}$Institute of Condensed Matter Theory and Optics, Friedrich-Schiller University Jena, 07743 Jena, Germany\\
$^{4}$Institute of Applied Physics, Abbe Center of Photonics, Friedrich-Schiller University Jena, 07745 Jena, Germany\\
$^{5}$Fraunhofer-Institute for Applied Optics and Precision Engineering IOF, 07745 Jena, Germany\\
$^{6}$Research Center for Materials Nanoarchitectonics, National Institute for Materials Science,  1-1 Namiki, Tsukuba 305-0044, Japan\\
$^{7}$Research Center for Electronic and Optical Materials, National Institute for Materials Science, 1-1 Namiki, Tsukuba 305-0044, Japan\\
$^{8}$Julius-Maximilians-Universität Würzburg, Physikalisches Institut and Würzburg-Dresden Cluster of Excellence ctd.qmat, Lehrstuhl für Technische Physik, Am Hubland, Würzburg, Germany\\
$^{9}$Fakultät für Physik, Munich Quantum Center, and Center for NanoScience (CeNS), Ludwig-Maximilians-Universität München, Geschwister-Scholl-Platz 1, 80539 München, Germany\\
$^{10}$Munich Center for Quantum Science and Technology (MCQST), Schellingstra\ss{}e 4, 80799 München, Germany\\
$^{11}$School of Mathematical and Physical Sciences, University of Sheffield, Sheffield S10 2TN, United Kingdom
}
\date{\today}

\begin{abstract}
Transition metal dichalcogenides represent a versatile platform to study strong light-matter interactions based on excitons and electrons in ordered lattices. Twist-engineering of moiré structures further enables the manipulation of the polaritonic nonlinearities via engineering the exciton landscape on the nanoscale. In this work, we demonstrate in-situ control of the optical saturation-based nonlinearity of moiré exciton-polaritons by phase space restriction via charge doping. Strong exciton-photon coupling is established in a gate-controllable MoTe$_2$-MoSe$_2$ heterobilayer, embedded in a spectrally-tunable open cavity. A small gate voltage can effectively lower the necessary polariton density by one order of magnitude to achieve a similar nonlinear saturation effect as in the charge-neutral case. Our microscopic description successfully explains the observed phenomena in the framework of Pauli blocking for the moiré superlattices with charge preoccupation.  
\end{abstract}

\maketitle


$Introduction-$Exciton-polaritons are composite bosonic quasiparticles formed by strong coupling between excitons in matter and photons in optical cavities \cite{PhysRevLett.69.3314}. These light–matter hybrids inherit a sizable optical nonlinearity from their excitonic component, which leads to distinct optical responses. Understanding the microscopic origins of nonlinearity is not only of fundamental importance but also enables promising applications, including all-optical logic gates \cite{ballarini2013all,timmer2026ultrafast,gc15-qsvf}, parametric amplifiers \cite{saba2001high,zhao2022nonlinear}, quantum light sources via polariton blockade \cite{PhysRevB.73.193306,munoz2019emergence,delteil2019towards}, quantum simulation and quantum computing \cite{o2009photonic,Kuriakose2022,kavokin2022polariton}. 

In conventional semiconductor microcavities, polaritonic nonlinearity arises from a combination of exchange interactions \cite{PhysRevB.58.7926,Shahnazaryan2017} and phase-space filling where Pauli blockade becomes significant at high exciton densities due to their underlying fermionic constituents \cite{PhysRevB.32.6601,PhysRevB.42.5147,PhysRevResearch.6.023033}. Additional nonlinear mechanisms include Rydberg blockade driven by long-range dipole–dipole interactions \cite{makhonin2024nonlinear}, and weaker Coulomb blockade in ultra confined systems \cite{PhysRevLett.79.1467,PhysRevB.73.193306}. However, realizing a substantial polariton energy shifts via these regimes usually requires high polariton density. Enhancing nonlinearity at low excitation density thus remains a central pursuit in polaritonics.

The advent of two-dimensional transition metal dichalcogenides (TMDC) has boosted this research field thanks to their giant exciton oscillator strengths \cite{PhysRevLett.113.076802,RevModPhys.90.021001}, simplicity in sample fabrication and ease of integration into optical microcavities \cite{schneider2018two}. TMDCs have thus emerged as a versatile platform for polaritonic nonlinearity studies. For example, trion-polariton or polaron-polariton in MoSe$_2$ monolayer (ML) \cite{emmanuele2020highly,PhysRevX.10.021011,PhysRevX.13.031036}, dipolaritons in MoS$_2$ homobilayer \cite{datta2022highly,louca2023interspecies,xiang2026electrically}, and Rydberg polaritons in WSe$_2$ \cite{gu2021enhanced} and WS$_2$ MLs \cite{shang2025chip} can exhibit considerable energy shifts with reduced pump flux. Recently, TMDC heterobilayers also opened a novel and unique avenue: by precisely controlling the twist angle, moiré superlattices can dramatically modify the excitonic landscape and, in turn, the strength of polaritonic nonlinearity through phase-space restriction \cite{zhang2021van,PhysRevResearch.6.023033}. Moreover, moiré superlattices enable electron-exciton interaction to manipulate the effective interactions \cite{tang2020simulation,shimazaki2020strongly,regan2020mott,wang2022light,campbell2022exciton,PhysRevLett.132.076902,scherzer2024correlated,kim2025moire}. However, a deeper investigation such as in-situ tuning of moiré-induced nonlinear interactions has remained elusive \cite{gr72-szwg}, which hinders the future on-chip applications. 

Here, we demonstrate robust and reversible electrostatic control of optical nonlinearity of moiré exciton-polaritons in a dual-gated MoTe$_2$-MoSe$_2$ heterobilayer. We interpret our experimental findings using a  microscopic theory of nonlinear phase space filling (NPSF) for moiré exciton-polaritons in the presence of charge occupancy of moiré sites. Our results establish a versatile electrical tunability of moiré polaritonic nonlinearity, and pave the way towards reconfigurable nonlinear nanophotonic applications.


\begin{figure}[t]
\includegraphics[width=\columnwidth]{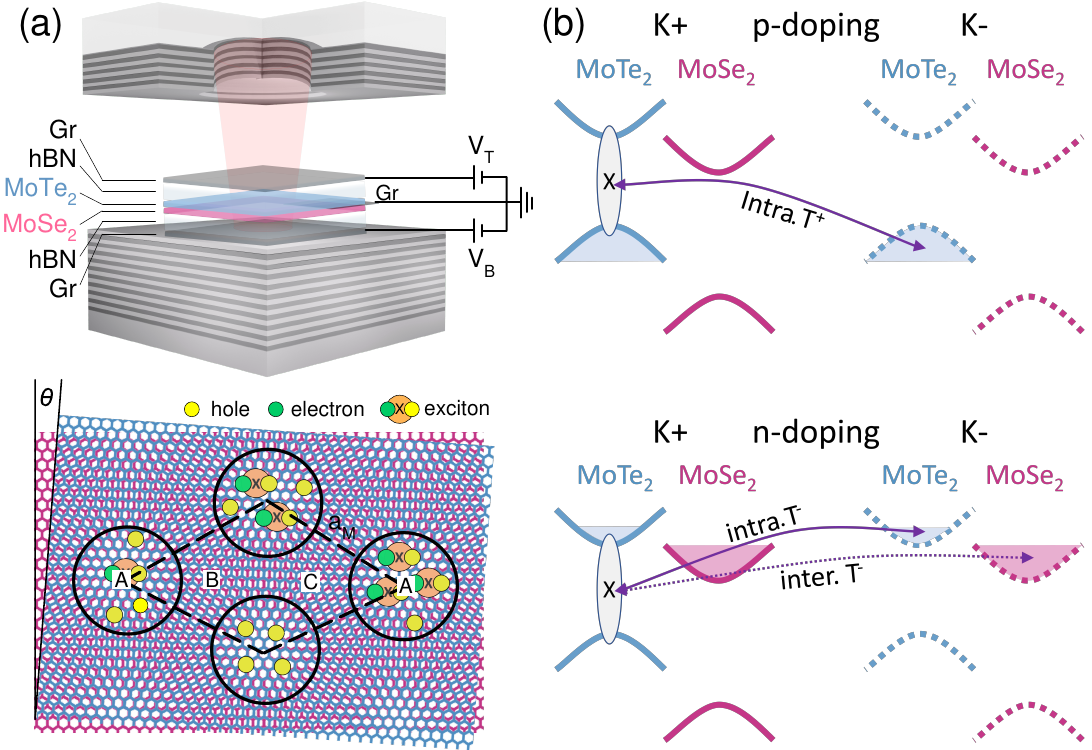}
\caption{\label{figure_1}(a) Top panel: schematics of open optical microcavity with a dual-gated MoTe$_2$-MoSe$_2$ HBL. Bottom panel: schematics of exciton (X) and charge filling into the high-symmetry point \textit{A} of the moiré superlattices for a R-type HBL. (b) Type-II band alignment in R-type MoTe$_2$-MoSe$_2$ HBL. Trion (T) states appear due to charge doping. Solid (dashed) lines represent spin-up (-down) bands.}
\end{figure}

\begin{figure*}[t]
\includegraphics[width=\textwidth]{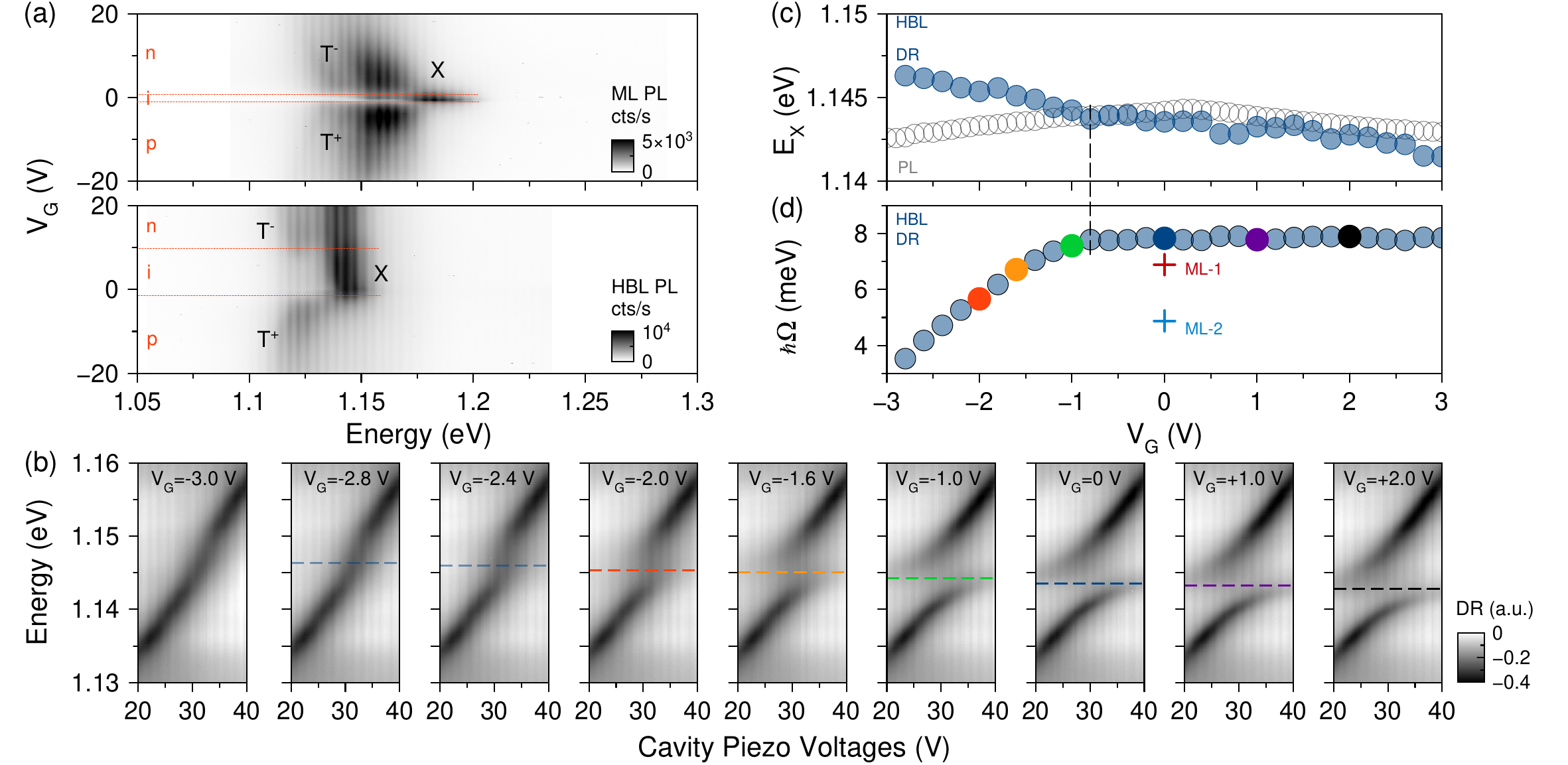}
\caption{\label{figure_2} (a) Gate tunable PL of MoTe$_2$ ML region (top) and MoTe$_2$-MoSe$_2$ HBL (bottom) of the charge tunable device, with negative (n), neutral (i) and positive (p) doping regimes. (b) Cavity detuning DR at different $V_G$. All graphs share the same color bar. The dashed lines mark moiré exciton resonances. (c) $V_G$-dependent exciton energies in HBL fitted from PL in (a) and DR in (b). (d) $V_G$-dependent Rabi splittings ($\hbar\Omega$) in HBL. Color symbols correspond to the $V_G$ applied in Fig. \ref{figure_3}(a). The Rabi gap energies of two ungated MoTe$_2$ MLs are shown by cross symbols.}
\end{figure*}

\begin{figure}[t]
\includegraphics[width=\columnwidth]{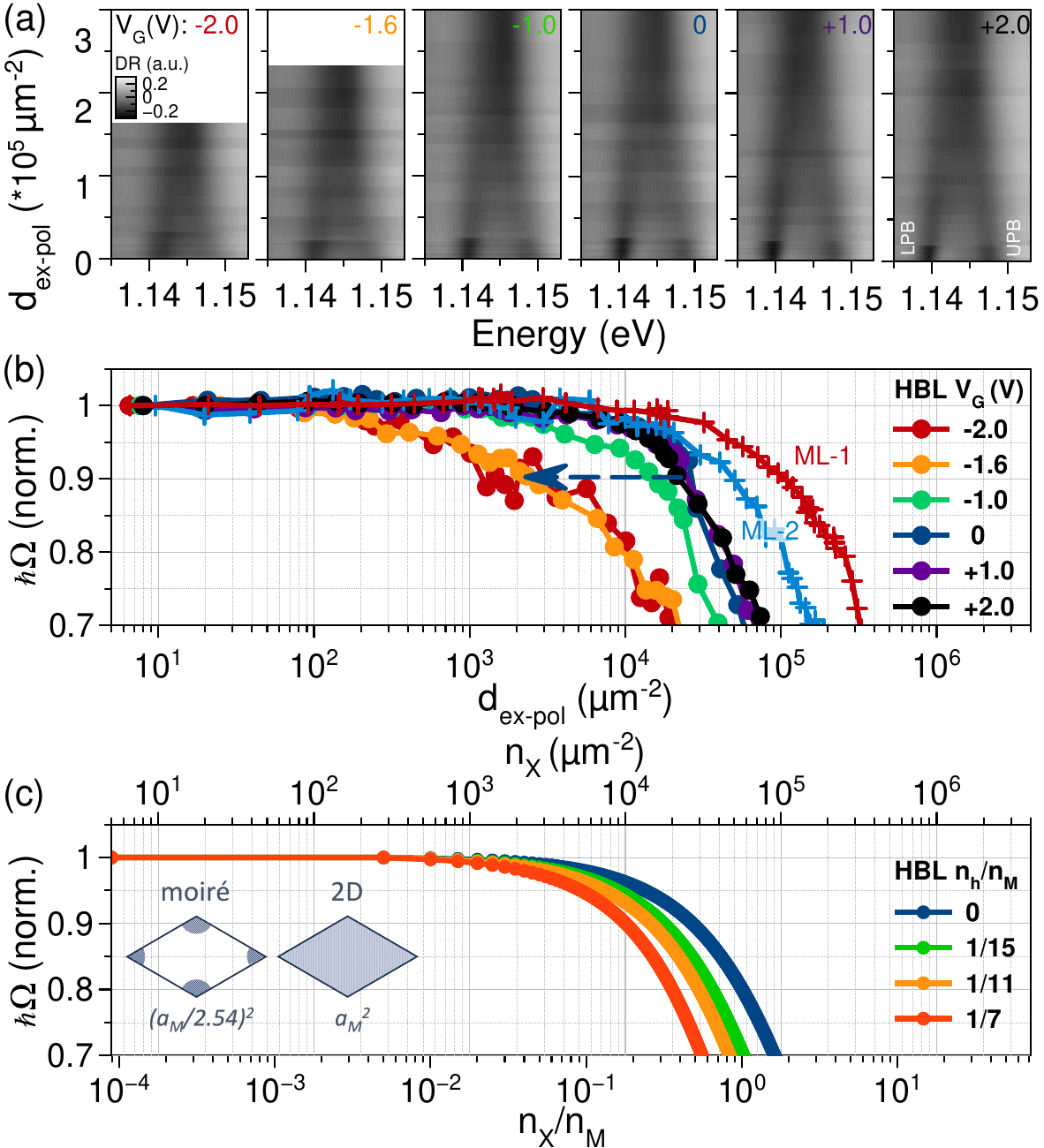}
\caption{\label{figure_3}(a) Density dependent zero-detuning DR spectra at different $V_G$ marked in Fig. \ref{figure_2}(d). The colorbar applies to all graphs. (b) Rabi gap saturation of gated HBL and ungated MoTe$_2$ MLs as a function of $d_{ex-pol}$. (c) Theoretical nonlinear saturation of moiré exciton-polaritons in HBL as a function of $n_X$ (top) and $n_X/n_M$ (bottom). Saturation curves with hole filling factors $n_h/n_M$ corresponding to  $\varepsilon_F=0,~9,~13,~21$ meV are presented. Inset: hole doping localized in a moiré cell $vs.$ delocalized in a 2D lattice.}
\end{figure}

\textit{Moiré~Exciton-Polaritons-}Figure~\ref{figure_1}(a) shows the schematics of our system corresponding to a van der Waals heterostructure placed inside a cryogenic open optical microcavity ($T = 3.5$~K). A voltage is applied to a piezo nano-positioner to drive the bottom distributed Bragg reflector (DBR), so that the cavity length and resonances can be finely tuned. Together with the flat bottom DBR, the concave top DBR (6~$\mu$m in diameter) confines discretized Laguerre-Gaussian modes. More details of these transverse cavity modes are in Ref. \cite{drawer2023monolayer,han2024situ,PhysRevLett.134.076902,han2025exciton}. 

We estimate the twist-angle of MoTe$_2$-MoSe$_2$ heterobilayer (HBL) to be $\theta$ $\sim$ $2^{\circ}$, yielding a moiré superlattice constant $a_M\approx$ 4.2 nm and a moiré density $n_M=1/a_M^2\approx5.6\times 10^4$ $\mu m^{-2}$ (Supplementary S1 \cite{supp}). The HBL is sandwiched by two hexagonal boron nitride (hBN) flakes with nearly identical thickness (Supplementary S3B \cite{supp}). The outermost graphene (Gr) layers serve as the top and bottom gates forming a plate capacitor, while the middle few-layer Gr acts as the contact gate. The entire heterostructure resides on the bottom DBR with pre-patterned gold electrodes for electrical contacts. 

Without the top mirror, we firstly measure gate voltage ($V_G$) dependent photoluminescence (PL) of the MoTe$_2$ ML and HBL regions in the charge tunable device. The top and bottom gate voltages ($V_T$ and $V_B$) are set to be same to $V_G$, while the contact gate is earth-grounded. All experimental methods are in Supplementary S1 \cite{supp}. The experimental results are compiled in Fig. \ref{figure_2}(a). MoTe$_2$ ML shows a typical ambipolar charging behavior, with symmetric exciton transitions to the positive and negative trion states. The charge neutrality at $V_G=-0.4$ V indicates an intrinsically weak electron doping. 

In comparison, the HBL maintains charge neutrality in a wider range of -2 V $\leqslant V_G\leqslant$ +10 V, with exciton PL emission redshifted by approximately 40~meV due to the typical moiré confinement effect. The hole doping effect below $V_G=-2$~V shows a complete exciton-trion crossover in stark contrast to the electron doping regime above $V_G=+10$~V. We note that previous reports found a regime in similar samples, where hybrid dipolar excitons with a similar charging behavior arise due to a resonant hybridization of the electron wavefunctions \cite{zhao2024hybrid}. However, since we observe no strong absorption feature from a higher-lying hybridized exciton, we believe that the detuning between the conduction band edges exceeds the electronic coupling strength. The charge tuning characteristics in Fig.~\ref{figure_2}(a) is thus strongly indicative of the type-II band alignment depicted in Fig.~\ref{figure_1}(b). 

Due to a large valence band offset of $\sim$300 meV, a positive trion state is formed by binding an extra hole in the spin-opposite valley within the MoTe$_2$ layer. Thus, the PL peak switches abruptly after the formation of positive trions. However, the negative trion states can form based on electrons within the same or different layer, generating intralayer or interlayer trions. Specifically, the formation of interlayer trion does not necessarily exclude the presence of excitons, such that coexisting PL signals of trionic and excitonic origins are possible for $V_G\geqslant+10$ V. Furthermore, the ordinary Landé g-factor as in MoTe$_2$ MLs \cite{PhysRevLett.134.076902} also agrees with a previous report for R-type MoTe$_2$-MoSe$_2$ HBL, suggesting highly intralayer contribution to the moiré excitons in our HBL [Fig.~S2(d) \cite{supp}]. We fit the PL spectra by Lorentzian function, and summarize $V_G$-dependent moiré exciton energies in Fig.~\ref{figure_2}(c). In the following, we concentrate only on the hole doping and neutral regimes in this HBL. 

Strong coupling of moiré excitons with discretized photonic modes is established in the fully assembled cavity. Figure \ref{figure_2}(b) shows clear anti-crossings in differential reflectivity (DR), featuring the formation of moiré exciton-polaritons [complete dateset and fit in Fig. S2(a) \cite{supp}]. As $V_G$ decreases to -3.0 V, we observe a progressive and reversible quenching of the Rabi gap between the upper and lower polariton branches (LPB, UPB) due to the reduction of exciton oscillator strength by the hole occupation of moiré pockets. 

By fitting the polaritonic modes with Lorentzian functions, we extract $V_G$-dependent exciton resonance energy and Rabi splitting, summarized in Figs. \ref{figure_2}(c) and \ref{figure_2}(d), respectively. We observe that the excitonic Stokes shift, extracted from the PL and DR results in Fig. \ref{figure_2}(c), continuously increases with electrostatic doping for $V_G\leqslant-0.8$ V. The Rabi splitting also monotonically decreases for $V_G\leqslant-0.8$ V, coinciding with the onset of Stokes shift (vertical line mark). We fabricate two additional ungated MoTe$_2$ MLs with similar hBN encapsulation as reference samples. Their strong coupling measurements are in Figs. S1 \cite{supp}, and the Rabi gap energies are marked by cross symbols in Fig. \ref{figure_2}(d).


\textit{Optical~Nonlinearity~of~Moiré~Exciton-Polaritons-}To probe the nonlinear optical response of moiré exciton-polaritons with charge filling in the superlattices, we study the optical saturation in density dependent DR measurements at six distinct $V_G$ corresponding to the color symbols in Fig. \ref{figure_2}(d). The occuring saturation of Rabi gaps in HBL are then compiled in Fig. \ref{figure_3}(a). The saturation behavior of ungated MoTe$_2$ MLs are in Fig. S1 \cite{supp}. To enable a more quantitative analysis, we convert the average pump power to polariton densities (Supplementary S2 \cite{supp}). We then fit and calculate the Rabi gap reductions in the gated HBL and ungated MoTe$_2$ MLs. 

To warrant comparability, the Rabi gaps are all normalized to the corresponding $V_G$-dependent values in Fig. \ref{figure_2}(d). Figure \ref{figure_3}(b) shows that the Rabi gaps of MoTe$_2$ MLs start to decrease at a polariton density $d_{ex-pol}\sim$$10^4$ $\mu m^{-2}$, in agreement with the excitonic Mott density in TMDC monolayers \cite{chernikov2015population}. Moreover, in charge neutral regimes, moiré exciton-polaritons in HBL show more pronounced saturation effects than the delocalized 2D exciton-polaritons in MoTe$_2$ MLs, since every moiré site cannot be occupied with an arbitrary amount of excitons. Specifically, as the initial Rabi gaps saturate by 10 \%, $d_{ex-pol}$ in HBL is 4 and 2.5 times smaller than those in MoTe$_2$ ML-1 and ML-2, respectively, even though the initial Rabi gap in HBL ($\hbar\Omega=7.8$ meV) is larger than that in ML-1 ($\hbar\Omega=6.9$ meV) and much larger than that in ML-2 ($\hbar\Omega=4.9$ meV) [Fig. \ref{figure_2}(d)]. 

Remarkably, we can further enhance the optical saturation effect in HBL by reducing $V_G$ below $-1$~V. Compared to the neutral regimes, the $d_{ex-pol}$ required to achieve a same saturation amplitude can be one order of magnitude smaller by applying a small gate voltage of $V_G=-1.6$~V [dashed arrow in Fig. \ref{figure_3}(b)]. It is also worth noting that, at the lowest exciton density, the Rabi gap at $V_G$ = $-1.6$ V is only $1.1$~meV smaller than that at charge neutrality ($\hbar\Omega\sim7.8$~meV for $V_G\geqslant -0.8$~V), and even exceeds the Rabi splitting in MoTe$_2$ ML-2.


\textit{Theoretical~modeling-}To investigate the role of NPSF due to Pauli blocking of moiré exciton-polariton, we firstly approximate the localized moiré exciton as
$x^\dagger_{\nu}
=
\int d\mathbf{r}_e\, d\mathbf{r}_h\,
\Psi_{\nu}(\mathbf{r}_e,\mathbf{r}_h)\,
a^\dagger_{\mathbf{r}_e} b_{\mathbf{r}_h}$,
where $\Psi_{\nu}$ is the exciton wavefunction ($\nu=0$ for the ground state), and $\mathbf{r}_e$ ($\mathbf{r}_h$) is the coordinate of electron (hole) in the 2D crystal. Owing to the composite nature of excitons, the operators satisfy
$[x_\mu,x^\dagger_{\nu}]
=
\delta_{\mu\nu}
-
D_{\mu\nu}$ \cite{combescot2008many}, with non-bosonic correction
\begin{align*}
D_{\mu\nu}
=
\int d\bar{\mathbf{r}}d\mathbf{r}d\mathbf{r}'
\Big[
&\Psi^\ast_{\mu}(\mathbf{r},\bar{\mathbf{r}})
\Psi_{\nu}(\mathbf{r}',\bar{\mathbf{r}})
a^\dagger_{\mathbf{r}'} a_{\mathbf{r}}\notag\\
&+
\Psi^\ast_{\mu}(\bar{\mathbf{r}},\mathbf{r})
\Psi_{\nu}(\bar{\mathbf{r}},\mathbf{r}')
b^\dagger_{\mathbf{r}'} b_{\mathbf{r}}
\Big].
\end{align*}

To capture qualitative behavior, we describe localized excitons using a suitable trial wavefunction,
\begin{equation}
\Psi_0(\mathbf{r}_e,\mathbf{r}_h)
\approx
(\pi L a_{\mathrm{X}})^{-1}
\exp\!\left[
-\frac{1}{2}
\left(
\frac{R^2}{L^2}
+
\frac{r^2}{a_{\mathrm{X}}^2}
\right)
\right],
\end{equation}
where
$\mathbf{R}=\frac{m_c}{M}\mathbf{r}_e+\frac{m_v}{M}\mathbf{r}_h$,
$\mathbf{r}=\mathbf{r}_e-\mathbf{r}_h$,
and $M=m_c+m_v$ with $m_c$ and $m_v$ being the conduction and valence band masses. The length $L$ is characterized by exciton center-of-mass motion localization and $a_X$ is characterized by the exciton size. Assuming coupling to a spatially uniform cavity mode, the density-dependent Rabi splitting is
\begin{equation}
\Omega_N
=
\Omega_0
\sqrt{
\mathcal{F}^{(N_s)}_{N+1}
/
\mathcal{F}^{(N_s)}_{N}
},
\label{eqn:OmegaN}
\end{equation}
where $\mathcal{F}^{(N_s)}_N$ is the normalization factor of the $N$-exciton wavefunction~\cite{combescot2008many,PhysRevB.73.115343,PhysRevResearch.6.023033} with
\begin{equation*}
\mathcal{F}^{(N{\mathrm{s}})}_N=
\sum_{m=1}^{N}
\frac{(-1)^{m-1}}{N}
\left[\frac{N!}{(N-m)!}\right]^2
\frac{\tilde{\sigma}_m}{N_{\mathrm{s}}^{m-1}}
\mathcal{F}^{(N_{\mathrm{s}})}_{N-m}.
\end{equation*}
Here, $N$ is the total number of excitons and $N_s$ is the number of moir\'e cells. The NPSF arising from $D_{\mu\nu}$ enter through
\begin{equation}
\tilde{\sigma}_n
=
\sigma_n
\exp\!\left(
-\frac{k_F^2 a_{\mathrm{X}}^2}{\chi_n}
\right),
\end{equation}
where $k_F$ is the Fermi momentum with local Fermi energy $\varepsilon_F=\hbar^2k_F^2/(2m_v)$ within a moir\'e cell. The coefficients $\sigma_n$ and $\chi_n$ are determined by exciton wavefunction $\Psi_0$, and depend on parameters $L$, $a_{\mathrm{X}}$, and band masses (Supplementary S3 \cite{supp}). We can then estimate theoretically the Rabi gap saturation by Eq.~\eqref{eqn:OmegaN}, and compare it with the experimentally observed $V_G$-dependent saturation behaviors in Fig. \ref{figure_3}(b). 

For the delocalized 2D case, even though the estimated hole density $n_h$$\approx$ 8.1$\times10^3$ $\mu$m$^{-2}$ at $V_G$ = $-1.6$ V (Supplementary S3B \cite{supp}) is nearly the same as the $d_{ex-pol}\sim10^4$ $\mu$m$^{-2}$ for the onset of optical saturation, $n_h$ is actually one order of magnitude smaller than the moiré density $n_M$. To account for this discrepancy, the inset of Fig.~\ref{figure_3}(c) illustrates the more localized hole doping in areas around the A point of a moiré cell \cite{zhao2024hybrid}, in striking contrast to the delocalized hole doping in 2D. The carrier density previously estimated for the 2D case is thus subject to a renormalization by assuming an effective moir\'e cell area of $(a_M/2.54)^2$ rather than $a_M^2$, reflecting the charge inhomogeneity across the moiré landscape. The effective hole density within moiré cells $n_h^\ast\approx2.54^2\times n_h\approx5.2\times10^4$ $\mu$m$^{-2}$ at $V_G=-1.6$ V thus agrees well with $n_M$, thereby accounting for the enhanced optical saturation. 

We further consider $a_X$ = 1.5~nm due to screening effect \cite{PhysRevB.102.115310}, $L$ = $2$~nm, and $m_c$ = $m_v$ = 0.6$m_0$ \cite{kormanyos2015k} with $m_0$ being the free electron mass. $\varepsilon_F=\pi \hbar^2 n_{h}^\ast/m\approx0-21$ meV can be tuned by varying the experimentally extracted hole filling factor $n_h/n_M$ from $0$ to $1/7$ [Fig. S4(a) \cite{supp}], where the latter marks the onset of trion instability \cite{Efimkin:PRB95-2017}. $\varepsilon_F=21$ meV also equals the trion binding energy $E_b$ at $V_G=-1.6$~V [Fig. S4(b) \cite{supp}]. The theoretical calculation in Fig. \ref{figure_3}(c) shows that as $n_h/n_M$ ($\varepsilon_F$) increases the saturation curves can well capture the qualitatively enhanced optical saturation effects. This suggests that the Pauli blockade mechanism indeed plays a crucial role in governing the nonlinear optical saturation of moiré exciton-polaritons, even though the moiré superlattices are only sparsely filled by the hole doping.


\textit{Conclusions-}We report the first experimental realization of in-situ electrical manipulation of optical nonlinearity of moiré exciton-polaritons. In a dual gated MoTe$_2$-MoSe$_2$ heterobilayer embedded in a cryogenic open optical microcavity, the localized hole doping in moiré superlattices via gate voltage can restrict the available phase space of moiré exciton-polaritons. As a result, to achieve a similar saturation effect, the necessary polariton density is lowered by one order of magnitude by applying a small gate voltage. Our theoretical model accounts for doping effects in the polaritonic Pauli blockade model for moiré cells, in qualitative agreement with the experimental observations. Our work underscores the highly efficient tunability of optical nonlinearities in moiré systems, and lay a solid foundation for further studies of charge-tunable dipolaritons with increased interactions and Bose-Hubbard polaritons in moiré superlattices with tunable on-site coupling. 


\begin{acknowledgments}

\textit{Acknowledgments-}C.S., B.H. and Z.L. acknowledge funding from the Deutsche Forschungsgemeinschaft (DFG) in the framework of SPP 2244 (funding number: Schn1376/14-2), DFG within the initiative for major equipment (Project INST184-234), and the Wissenschaftsraum ElLiKo funded by the MWK via the Volkswagen foundation. 
C.S. Acknowledges support by the European Research Commission (ERC) within the project Dual-Twist (Grant agreement 101170213).
B.H. acknowledges the Alexander von Humboldt-Stiftung for the fellowship grant, and supportings from National Natural Science Foundation of China (NSFC) under Grant No.~12304012. 
M.E. acknowledges funding from the Carl von Ossietzky Universität Oldenburg through a Carl von Ossietzky Young Researchers' Fellowship. 
M.S.R and C.C. acknowledge financial support from the Ministry for Culture and Research of the Lower Saxony State “Professorinnen für Niedersachsen” and from the German Federal Ministry of Education and Research (Professorinnenprogramm III), the Wissenschaftsraum ElLiKo funded by the MWK via the Volkswagen foundation, and the fundings from DFG project numbers 398816777 (SFB 1375, project A8) and 547611111 (WHAT-A-TWIST).
F.E. acknowledges support by DFG (SFB 1375, project B6), and BMBF FKZs 16KISQ087K and 13XP5053A. 
K.W. and T.T. acknowledge support from the CREST (JPMJCR24A5), JST and World Premier International Research Center Initiative (WPI), MEXT, Japan. 
K.W.S. is supported by Xiamen University Malaysia Research Fund (Grant No. XMUMRF/2025-C15/IPHY/0005). 
O.K. acknowledges the support from UK EPSRC grants EP/Y021339/1 and EP/X017222/1.
A.H. acknowledges funding by the Deutsche Forschungsgemeinschaft (DFG, German Research Foundation) within the Priority Programme SPP 2244 2DMP and the Germany's Excellence Strategy under grant No. EXC-2111-390814868 (MCQST).
A.P., S.S., S.B. and S.H. acknowledge financial support from the Würzburg-Dresden Cluster of Excellence on Complexity, Topology and Dynamics in Quantum Matter ctd.qmat (EXC 2147, DFG project ID 390858490).

\textit{Author~Contributions-}B.H. and C.S. conceptualized and supervised the work. K.W. and T.T. synthesized the hBN crystals. Z.L. and N.N. fabricated the dual-gated samples. F.E. and M.E. prepared silica mesa. A.P., S.S., S.B. and S.H. sputtered DBR on the mesa. V.S. conducted the AFM measurements. M.S.R. and C.C. calculated the refractive index of MoTe$_2$ monolayer. J.D. and B.H. developed codes for data acquisition. B.H. and Z.L. conducted the experiments. B.H. processed the experimental data. B.H., C.S., A.H., Z.L., K.W. and O.K. contributed to data interpretation. K.W.S. and O.K. provided theoretical modeling. B.H. and K.W.S. wrote the manuscript with inputs from all authors. 

\textit{Supplementary Information-}contains Refs. \cite{banu2023first,ohtake2022structure,PhysRev.136.B864,RevModPhys.74.601,PhysRevB.54.11169,PhysRev.140.A1133,PhysRevB.50.17953,PhysRevLett.77.3865,grimme2010consistent,PhysRevB.74.035101,PhysRev.84.1232}.
  
\end{acknowledgments}

\bibliography{apssamp}

\end{document}